# Ferroelectric chiral nematic liquid crystals: New photonic materials with multiple bandgaps controllable by low electric fields

J. Ortega, C.L. Folcia, J. Etxebarria

*Department of Physics, Faculty of Science and Technology, UPV/EHU, Bilbao, Spain*

T. Sierra

*Instituto de Nanociencia y Materiales de Aragón (INMA), Química Orgánica, Facultad de Ciencias. CSIC-Universidad de Zaragoza, Pedro Cerbuna 12, ES-50009 Zaragoza, Spain*

e-mail address: j.etxeba@ehu.es

# Ferroelectric chiral nematic liquid crystals: New photonic materials with multiple bandgaps controllable by low electric fields

We have carried out a spectroscopic study on a prototype ferroelectric nematic material (RM7343) doped with a non-polar chiral compound. The mixture presents two chiral nematic phases, one conventional (N*) and the other polar ($N_F$*), whose behaviours under electric fields $E$ are totally different. On the one hand, the N* phase shows a single reflection band if $E$ is not very high, while fields perpendicular to the helix of a few V/mm are already sufficient to produce multiple bands in the $N_F$* phase. On the other hand, the pitch of the $N_F$* phase grows notably on increasing fields, however remaining practically constant in the N* phase. Both effects have been explained in terms of the different type of interaction with $E$ in each phase (ferroelectric in $N_F$* and dielectric in N*). We argue that the $N_F$* phase behaves as a photonic material with multiple gaps tunable by small fields, which presents important potentialities for applications. A study of the bandgaps has been carried out through the analysis of the dispersion relation of the optical eigenmodes in the $N_F$* phase under field. Finally, an example of the potentialities of the $N_F$* phase within the field of nonlinear optics is briefly presented.

Keywords: ferroelectric nematic phase; chirality; polar order parameter; multiple gap photonic materials, nonlinear optics.

## 1. Introduction

The discovery of the ferroelectric nematic ($N_F$) phase has given a great boost to the research of the basic science of liquid crystals (LCs) [1-13]. In the new phase, the head-to-tail molecular symmetry typical of traditional LCs is broken and this gives rise to the appearance of a polarization $P$ along the molecular director **n**. Usually the $N_F$ phase is obtained by cooling the ordinary nematic phase N, although it can also appear directly from the isotropic liquid [14]. The most striking characteristics of the $N_F$ phase are the existence of high $P$ values (of the order of $\mu C/cm^2$) [3,8], giant dielectric constants (of the order of $10^4$) [15] and a high non-linear optical susceptibility (of the order of 10 pm/V) [16]. All these characteristics, together with the fluidity of an ordinary nematic, allows us to think of these compounds as authentic revolutionary materials that can generate real breakthroughs and innovations in LC-based technologies.

Very recently, the consequences of the introduction of chirality in the molecules that give rise to the $N_F$ phase have begun to be explored [17-20]. It has been found that the resulting structure presents a twist similar to that of ordinary cholesterics with a well-defined helical pitch $p$, but presenting locally polar order (see Fig. 1, left). This new phase is denoted $N_F^*$. Materials exhibiting this phase are obtained by adding a small proportion of (non-polar) chiral component to $N_F$ prototype materials (RM734 or DIO) [1717,18], or by slight chiral modifications in the chemical structure of those prototype molecules [19]. Interestingly, it has been found that selective reflection in materials with the $N_F^*$ phase can be reversibly tuned by small electric fields perpendicular to the helix axis, in great contrast to the behaviour observed for ordinary N*, where the electric fields must be much higher and gives rise to very poor tuning. For example, in ref. [17] it is reported that fields of the order of 100 V /mm produce shifts of about 100 nm in the colour of the selective reflection in a mixture of RM734 and a chiral dopant, with optical pitch in the visible range. This effect is truly outstanding and, in fact, the phenomenon has been considered as the "holy grail of the LC display applications", due to the great possibilities it offers for the construction of many practical devices. Therefore, it seems of interest to continue the research on the $N_F^*$ phase in new materials.

In this work we present a characterization of the spectroscopic properties of a $N_F^*$ material as a function of the electric field and temperature. We have confirmed the tuning of the selective reflection band with electric fields even smaller than previously reported (of the order of 1 V/mm). In addition, and in agreement with previous works [18], we have found evidence of the existence of the so-called full-pitch optical band, which occurs at a wavelength $\lambda=2pn$, where $n$ is an average refractive index (that is, the wavelength is about twice that of the ordinary selective reflection band). The full-pitch band is also accompanied by higher-order harmonics, whose sizes and positions can be controlled by the electric field. We thus argue that $N_F^*$ phases are new photonic materials with multiple band gaps and electric field tunability. This fact further widens the range of applicability of the $N_F^*$ phase. As an example, we also present here experimental evidence of nonlinear optical effects related to the complex structure of the photonic band gaps.

This paper is organized as follows. In section 2 we will present the studied material and experimental equipment. The spectroscopic results will be shown in section 3 and discussed in section 4, where we will analyse the dispersion relation of the optical eigenmodes in the $N_F^*$ phase, explicitly showing the appearance of multiple gaps when an electric field is applied. Finally, in section 5, we will show some optical second-harmonic generation (SHG) measurements that point to the possibility of new potentialities of these materials in the field of nonlinear optics based on the existence of multiple gaps in them.

**2. Experimental**

The material studied was a mixture of RM734 and the non-polar chiral dopant D* (compound 2 of reference [21]) in a proportion of 5.8% wt. The sample was a plane-parallel cell made of two glasses treated with polyvinyl alcohol and parallel rubbed. Two ITO strip electrodes separated by 5 mm on one of the cell glasses allowed for the application of in-plane electric fields parallel or antiparallel to the rubbing direction. The cell thickness was 15 μm. The phase sequence on cooling was the following:

Isotropic-165 ºC-N*-116 ºC-N$_F$*-97 ºC-Crystal.

The material was introduced by capillarity into the cell. Initially, the alignment was not very good in the N* and N$_F$* phases, but it could be substantially improved by prolonged application of low-frequency AC voltages (around 1 Hz) with amplitudes of several tens of volts in the N$_F$* phase. The measurements were carried out on samples that presented a typical cholesteric texture with a few oily streaks, indicating a fairly reasonable alignment (Fig.1 right).

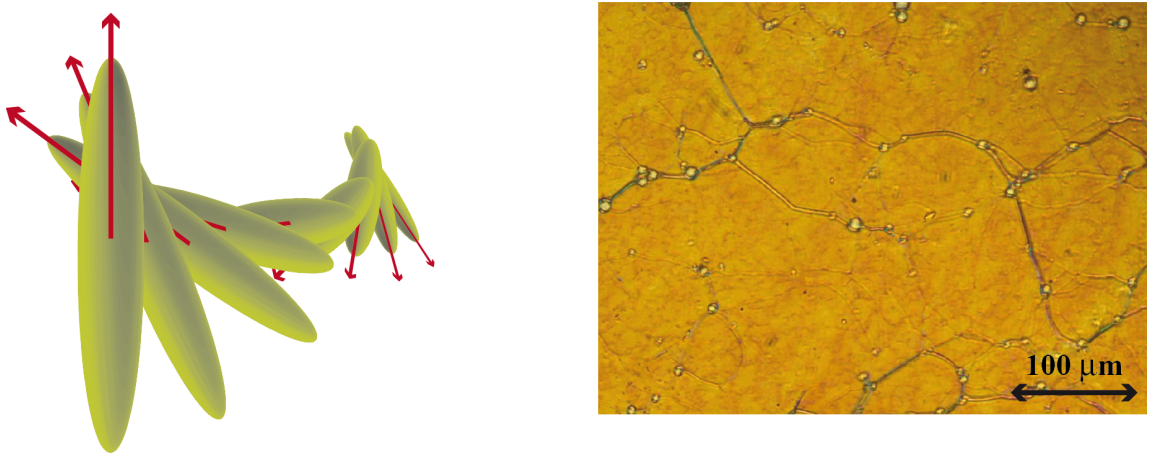

**Fig. 1**. Left: Scheme of the molecular arrangement in the N$_F$* phase. Red arrows represent the local polarization. Right: Texture of the N$_F$* phase at 105 ºC under crossed polarisers. The N* phase has a similar texture but with a different colour.

Reflectance spectroscopic measurements were performed with a fiber-optic spectrometer (Avantes) using natural light from a deuterium-halogen lamp. For the SHG measurements, a Q-switched pulsed Nd:YAG laser operating at a wavelength of 1.064 μm was used. The beam incident on the sample was collimated and had a diameter of 1 mm. The light intensity of the second harmonic was detected with a photomultiplier. Details of the experimental setup can be found in ref. [22]

3. Results

Figure 2a shows the reflection spectra of the material at several selected temperatures in the N* and $N_F$* phases, and Fig. 2b shows the temperature dependence of the pitch *p* deduced from the spectra. It was assumed that the wavelength of the centre of the band (half-pitch band) is *λ=pn*, where *n* is an average index *n*=1.8. This value was obtained from measurements in the isotropic phase using an interferometric method. The significant molecular rearrangement at the N*-$N_F$* transition shows remarkable pitch changes in both cholesteric phases. Near the transition, in the N* side, there is a rapid variation of the pitch length. In the $N_F$* the pitch is longer, which can be a consequence of the tendency of the dipoles to arrange parallel to each other. This fact results in a decrease of the effective helical twisting power of the chiral dopant in the ferroelectric phase.

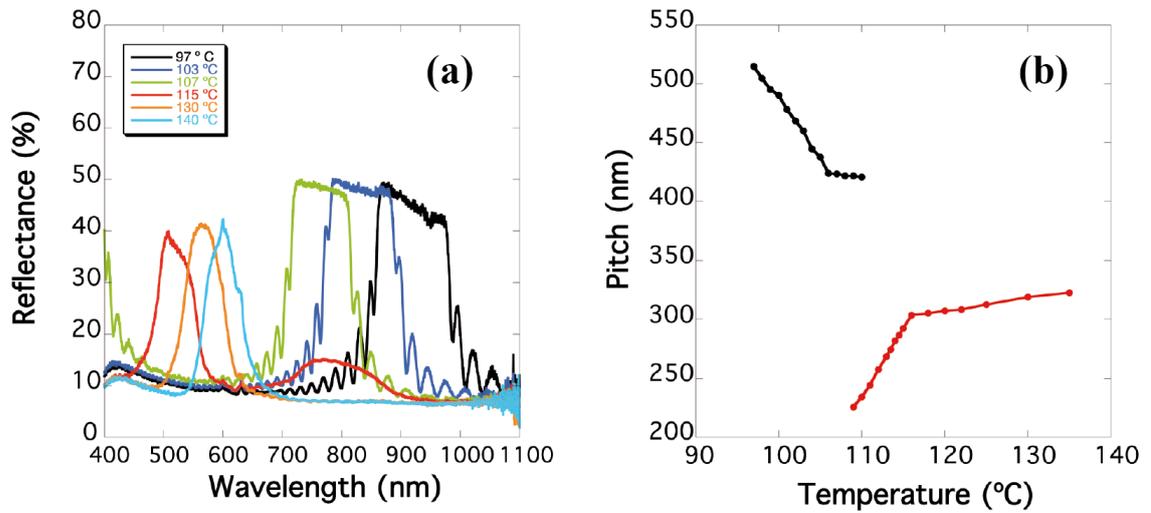

**Fig. 2.** a) Reflectance spectra of the sample at different selected temperatures. b) Pitch length vs. temperature (N* and $N_F$* phases are represented by red and black lines respectively). A phase coexistence range can be observed.

More interesting is the behaviour observed in the $N_F$* phase under electric fields. As can be seen in Fig. 3a, the application of small DC voltages produces, in addition to the red shift of the selective reflection band, the appearance of new reflectance bands at shorter wavelengths. In reference [18] a similar behaviour was reported, with the appearance of two new peaks at both sides of the main band (or half-pitch band). In our case, the spectral range of the spectrometer used prevented the detection of the band to the right of the half-pitch band. However, our measurements provide indirect evidence of the existence of a full-pitch–band. Indeed, as the field increases, peaks at shorter wavelengths are reinforced and their number grows. We have been able to detect up to 3 new peaks (at 0.671, 0.508 and 0.417 μm for an electric field $E$=3.6 V/mm, see Fig. 3b). These new bands occur at wavelengths given by $2pn/m$ with $m$=3,4,5, and are explained as successive harmonics (third, fourth and fifth) of the full-pitch band. The fact that the wavelengths of the high-order peaks do not coincide with exact submultiples of that of the full-pitch band is due to the dispersion of the mean refractive index on the wavelength. Finally, for larger fields all the bands disappear.

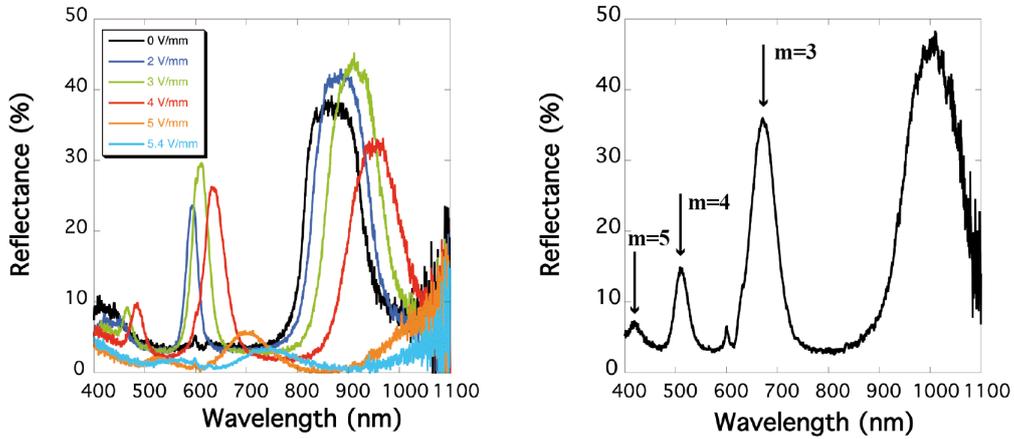

**Fig. 3.** a) Reflectance for different applied electric fields at 104 ºC. b) Detail of the different harmonics of the full-pitch band gap (E= 3.6 V/mm)

4. Discussion

As is well known, in the conventional N* phase the head-to-tail symmetry of the nematic director gives rise to a helical structure with a periodicity equal to $p/2$. Consequently, at normal incidence, Bragg's law predicts a single selective reflection at $\lambda=2n\ (p/2)=pn$. In contrast, in the $N_F$* phase the head-to-tail symmetry does not exist any longer, and the true structural periodicity is $p$. However, this does not mean that all the optical properties of the $N_F$* phase are substantially modified, because the optical indicatrix maintains the $p/2$ periodicity in the absence of applied fields. In fact, at zero field the $N_F$* phase only shows one selective reflection band at $\lambda=pn$ in the usual way.

The differences appear when an electric field $E$ is applied. In the N* phase the interaction is of dielectric type and it is quadratic in $E$. As a consequence, twist torques induced by the field are, in general, much smaller in the N* than in the $N_F^*$ phase and cannot compete against typical surface anchoring forces. Therefore, the deformation of the helix keeps the periodicity of the system constant (equal to *p*/2) until the field is strong enough to unwind the helix abruptly. This gives rise to a reflectance under field where the half-pitch band is still the fundamental reflection, with the possibility of appearing additional small contributions at shorter wavelengths *pn*/*m* (*m*=2,3,...), which can only be observed for very intense fields before unwinding the helix [23,24]. Such high order reflections simply reflect the contribution of certain Fourier components in the helix distortion. A typical profile of the molecular azimuth as a function of the space coordinate along the helix (*z*) is depicted in Fig. 4a.

In the $N_F^*$ phase the main interaction is ferroelectric and, therefore, linear in *E*. This interaction produces a totally different structural distortion: the nematic director is reoriented according to the polarity of the field, producing an azimuthal profile like the one indicated in Fig. 4b, where the periodicity is clearly the full pitch *p*. Therefore, a corresponding reflection band at *λ*=2*pn* must appear. Likewise, the high frequency bands appear with higher or lower strengths depending on the specific shape of the distortion and the relative magnitude of their Fourier components.

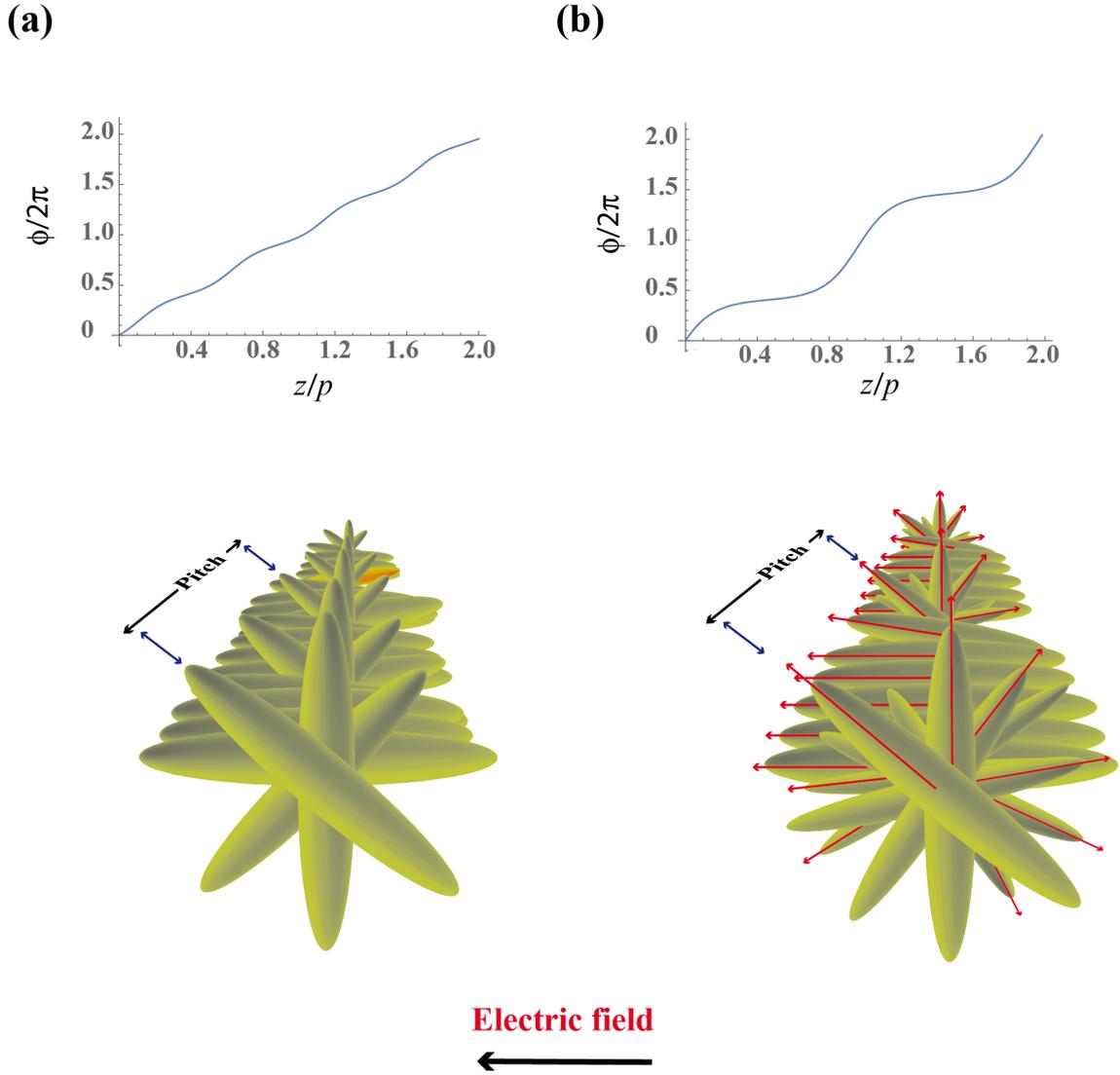

**Fig. 4.** Top: Azimuth profiles of the molecular director vs. $z/p$ for the N* (a) and $N_F$* (b) phases under strong distortion induced by an electric field perpendicular to the helix axis. For the simulation a quadrupolar coupling with the electric field has been assumed in the N* phase. The electric field values are 4300 V/mm (for a dielectric anisotropy $\Delta\varepsilon$=11.6) and 6.1 V/mm for the N* and $N_F$* phases respectively. Bottom: Schematic representation of the director distortion for the corresponding phases.

On the other hand, the fact that $p$ grows with $E$ is explained as due to the fact that an enlargement of $p$ compensates the increase of twist elastic energy in the regions where the molecular orientation changes rapidly (see Fig. 4b top) [17]. This increase in $p$ then gives rise to a redshift with $E$ in all reflection bands.

From a more quantitative point of view we will now try to find an approximate expression for the director profile as a function of $E$ in the $N_F^*$ phase. The azimuth angle of the director $\phi(z)$ with respect to a reference direction perpendicular to the helix (rubbing axis) verifies the differential equation [25]:

$$K_2 \, d^2\phi/dz^2 = PE \sin\phi, \qquad (1)$$

where $K_2$ is the twist elastic constant. In equation (1) the dielectric and flexoelectric terms have been omitted, keeping only the ferroelectric interaction.

To obtain $\phi(z)$ we must solve (1) with the boundary conditions:

$$K_2 \, [d\phi/dz \, (0) - 2\pi/p_0] - W \sin\phi(0) = 0$$

$$K_2 \, [d\phi/dz \, (d) - 2\pi/p_0] + W \sin\phi(d) = 0, \qquad (2)$$

where $d$ is the thickness of the sample and $p_0$ is the pitch in the absence of field. Furthermore, we have assumed an anchoring energy per unit area given by $-W \cos\phi$, where $W$ is a constant. Conditions (2) contain as special cases the situations of strong anchoring ($\phi(0) = \phi(d) = 0$) if the term in $W$ is dominant, or totally free anchoring ($d\phi/dz \, (0) = 2\pi/p_0$, $d\phi/dz \, (d) = 2\pi/p_0$) if $W$ is very small.

A typical solution of (1) can be seen in Fig. 4b. We have taken as parameters $K_2 = 6.5$ pN, $P = 0.045$ C/m$^2$, $p_0 = 0.488$ μm, $d = 10$ μm, and $W = 1 \times 10^{-6}$ J/m$^2$. Fig. 5 shows the reflectance of the sample for $\phi(z)$ profiles under different electric fields. The calculations were carried out using the Berreman method for eigenstates of light polarization, i.e., light modes that retain the polarization state after traversing the sample. The diffracting features of the system are mainly present in one of the eigenstates which is the one depicted in the figure. The simulation presents a good agreement with the experiments: red shifts for all reflection bands on increasing *E* (Fig. 5a) and appearance of high-order bands for small electric fields (several V/mm) (Fig. 5b).

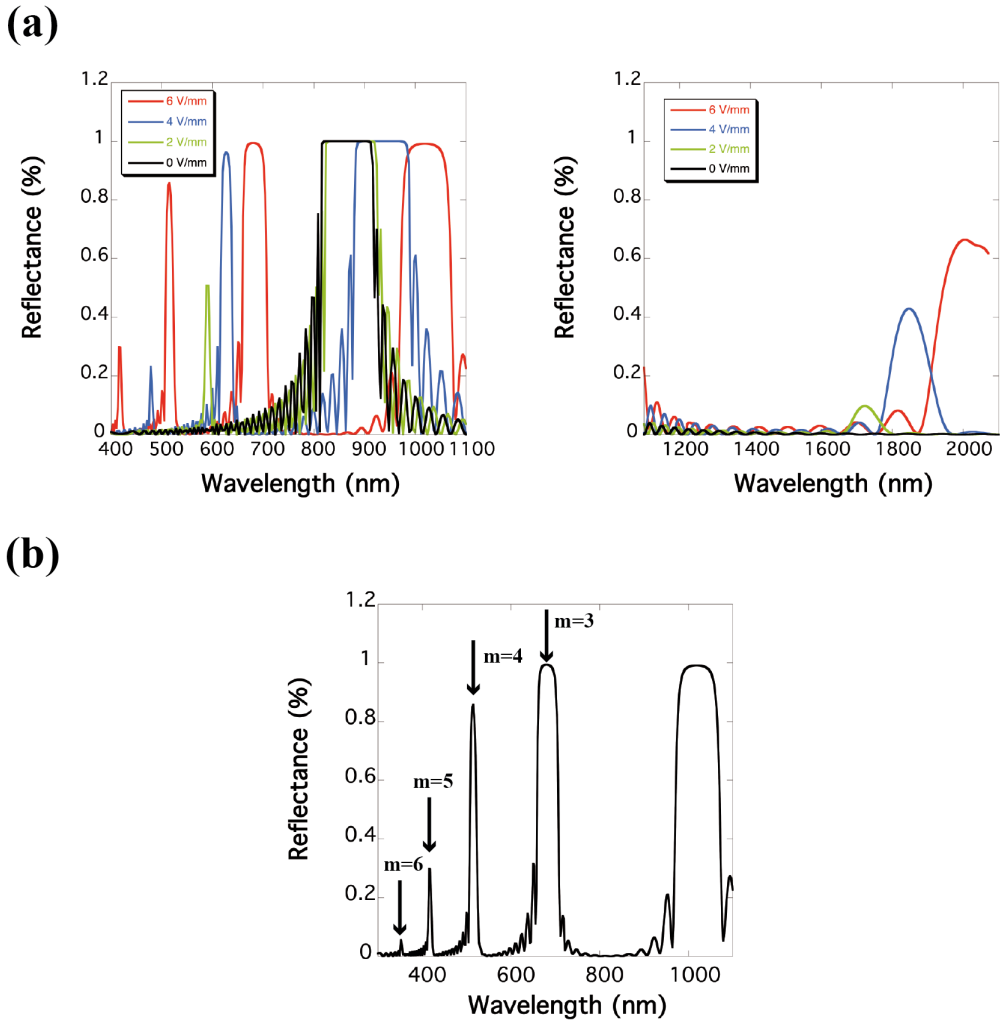

**Fig. 5.** a) Reflectance spectra of the diffracting eigenmodes calculated for a sample of 10 μm thickness for different electric fields. The two figures on the top represent different regions of the spectrum. b) Detail of the different reflection bands under strong distortion of the helical structure (*E*=6 V/mm).

It is now interesting to calculate the dispersion relations of the optical eigenmodes for the obtained director profiles. A similar work was carried out some time ago for light propagation along the helix axis of a normal N* distorted under field [26].

The oblique incidence has also been treated in the literature [27,28] for non-distorted N* and smectic C* phases. Here, we will just consider the case of incidence along the helix in a $N_F$* structure distorted by a small field. Given a frequency $\omega$, the eigenmodes and the dispersion relations $\omega(k)$ can be obtained by diagonalizing the so-called propagation matrix $F(\omega,z,p)$, which is a 4x4 matrix relating the *x* and *y* components of the electric and magnetic fields at *z* and *z+p*. For each frequency, the matrix $F(\omega,z,p)$ is routinely obtained with the Berreman method, and its eigenvalues *v* give the 4 corresponding wave vectors *k* through the expression *v*=exp(*ikp*). There are 2 forward modes and 2 backward modes that are identical to each other except for the direction of propagation. Figure 6 shows the dispersion relations for *E*=0 (a) and *E*≠0 (b) drawn within the first Brillouin zone of the structure, that is, in the interval ($-\pi/p, +\pi/p$). This representation is known as the reduced zone scheme. Only the modes that propagate without attenuation (*k* real) are shown. The ordinary and extraordinary refractive indices were assumed to have a small dependence with the wavelength, which was modelled with a Sellmeier-type formula.

For *E*=0, we obviously get the well-known dispersion of the N* phase, since N* and $N_F$* are optically equivalent. There is a gap at the centre of the zone (*k*=0) in one of the dispersion branches (red curve in Fig. 6a) whose polarization is predominantly circular with the same handedness as the helix (except near the edges of the gap and at very high frequencies). The other branch does not present any gap (black curve in Fig. 6a) and the polarization of the eigenmodes is essentially circular with opposite handedness to that of the helix (except at very high frequencies). The situation changes if *E*≠0 (Fig. 6b). As can be seen, even a small field (typically E ~1 V/mm) gives rise to gaps. Gaps (numbered as 1-4 in Figs. 6b,c) are opened for the two dispersion branches both near the centre (*k*=0) and near the edges of the Brillouin zone ($k=\pm\pi/p$). The lowest frequency gap (number 1 in Fig. 6c) corresponds to the full-pitch band, and appears at a frequency approximately equal to half that of the gap in the absence of field. The successive gaps (numbers 2-4 in Fig. 6c) have frequencies approximately multiples of the full-pitch gap. The new gaps are especially wide and strong (have larger attenuation coefficients and, therefore, lead to higher reflectances) for polarizations with the same handedness as the helix (red curve). There are intervals of frequency of selective reflection, where a gap is produced only for one of the two polarizations and regions where reflection gaps occur for both polarizations (see for example the central zone of gaps 1 and 2 in Fig 6c). In general, the gaps move and become wider on increasing $E$. Moreover, their frequency ranges further change, since the periodicity $p$ is also dependent on $E$. In other words, the $N_F^*$ phase can be considered as a multigap photonic material with tunability under electric field.

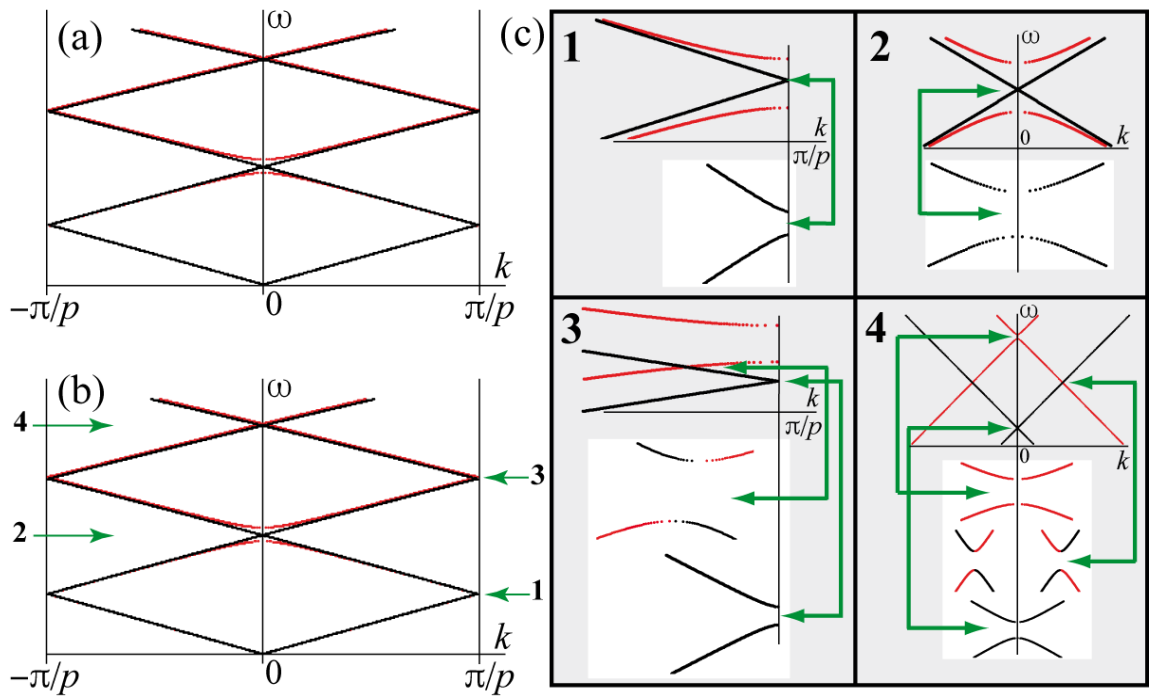

**Fig. 6.** (a) Dispersion relation $\omega(k)$ represented in the reduced Brillouin zone scheme for the $N_F^*$ structure under zero field. A single gap results for $k=0$ in one of the branches.

(b) $\omega(k)$ for the $N_F^*$ phase when the structure is distorted by a small electric field perpendicular to the helix axis. There are gaps close to the zone centre (2 and 4) and to the zone edges (1 and 3). Regions 1 to 4 are detailed in (c) where some $\omega$ intervals are further enlarged at different scales (white rectangles).

## 5. SHG measurements

As an example of the potentialities that materials with multiple gaps can display, we now present some SHG measurements in which an enhancement of the SHG signal has been detected due to a resonance effect between gaps.

It has been well known for years that some nonlinear photonic materials can allow efficient SHG [29, 30]. The SHG signal is enhanced because the photonic gaps produce resonance effects both on the fundamental and SHG waves. This is the basis of the operation of certain artificial multilayer structures specifically manufactured to present high conversion efficiency for the SHG process. However, the fabrication of these structures is difficult and expensive, and it is evident that self-assembled materials such as LCs can offer great advantages at this point.

In relation to this question, Yoo et al [31] observed years ago the existence of a special type of phase matching for SHG in the helical structure of a smectic C* ferroelectric LC using two fundamental waves counterpropagating along the helix when the wave SHG is close to one of the edges of the selective reflection band. However, the experimental geometry is complicated by the requirement to have two counter waves. In addition, the non-linear susceptibilities of the smectic C* materials are small [32], there is only a single photonic gap in them, and the final result was not extraordinary.

Although the detailed theory is complicated [33], it is intuitive to understand the possibility of attaining very effective phase matching in multigap materials when both the fundamental and SHG beams are near the edges of two photonic bandgaps. In this case, the group velocities for both waves are very small and the electric fields within the cavity undergo high amplifications. In fact, it can be shown that this situation gives rise to an important enhancement of the SHG especially if there is a strong overlap of both fundamental and SHG modes [34].

We have examined the possibility of this phase matching in our material. In view of the specific wavelengths of the observed reflectance peaks (Fig. 3), the candidate photonic gaps to carry out the conversion were the half-pitch band (with a wavelength close to 1.064 μm) and the fourth harmonic of the full-pitch band (wavelength near 0.532 μm). The experimental geometry is simpler here than in the special phase matching case of the two counterpropagating waves, since only one input beam is necessary. Fig. 7a shows the results obtained when the 1.064 μm input wave was linearly polarized along the applied electric field. Two maxima can be observed in the SHG signal at $E$=3.6 and 5.4 V/mm. We interpret these maxima as due to conversions of 2 photons of wavelength 1.064 μm near the half-pitch band into one photon of wavelength 0.532 μm near the fourth band. The observation of 2 peaks is understood in terms of resonances between the 2 long-wavelength edges and the 2 short-wavelength edges of the bands, respectively (see Figs. 7b and c).

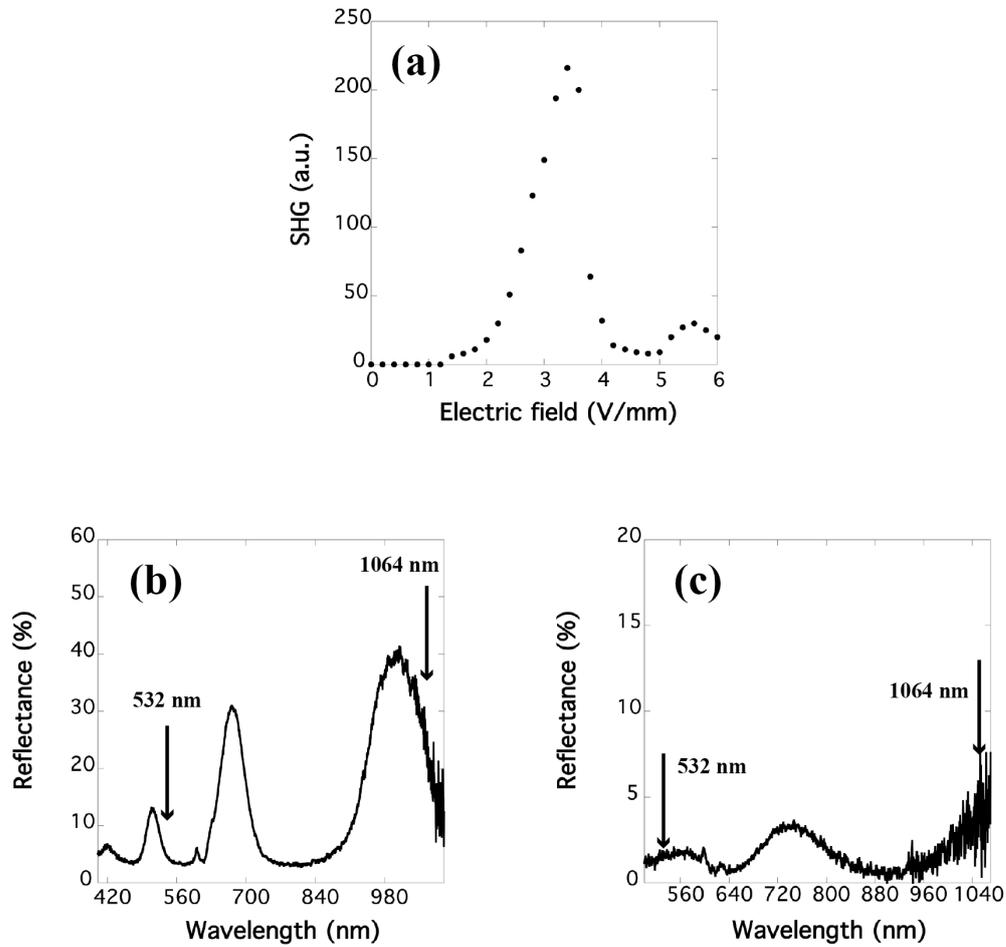

**Fig. 7.** SHG signal as a function of the applied electric field. The fundamental input light was polarized along the electric field. (b) Reflectance of the sample for the electric field corresponding to the main maximum of the SHG signal (3.6 V/mm). Both fundamental and second-harmonic wavelengths coincide approximately with the long-wavelength edges of the second and fourth photonic bands simultaneously. (c) Reflectance spectrum at 5.4 V/mm corresponding to the secondary SHG maximum. In this case the fundamental and SHG wavelengths are roughly positioned at the short-wavelength edges of the second and fourth harmonics respectively.

Though conceptually the helix-assisted conversion has been demonstrated, the efficiency of the process should be improved to be useful, since the SHG signal is still higher when the helix is suppressed and the input field is parallel to **n**. There are at least two reasons to explain why the performance is not as good as expected. First, the two wavelengths 1.064 and 0.532 μm do not coincide simultaneously with the edges of the second and fourth-order gaps in any case. Second, it is evident that in a photonic material the efficiency is greatly affected by the quality factor of the resonant cavities that come into play, and most probably this quality is quite poor in our case (especially for a cavity of an order as high as 4). If lower order gaps are wanted to be used for the same conversion process (full and half-pitch bands for fundamental and SHG wavelengths respectively), the $N_F^*$ materials must have shorter pitches, and this would require to increase the concentration of the chiral component in our mixture. Unfortunately, RM734 does not give homogeneous mixtures with higher concentration of compound D* than the one used in the present work.

## 6. Conclusions

We have carried out a spectroscopic study of the new helical phase $N_F^*$ using reflectance measurements and analysing the dispersion relations as a function of the applied field. In agreement with previous works, we have found that the application of low voltages produces important variations of the structural periodicity *p* and gives rise to the appearance of a rich scheme of photonic gaps in the centre and edge of the Brillouin zone. We then conclude that the $N_F^*$ phase behaves as a material with multiple tunable gaps under field. These features are very interesting and can be applied in many fields such as LC displays, shutters, tunable lasers, or new possibilities in non-linear optics. To explore some of these potentialities, it would be desirable to have $N_F^*$ materials with shorter pitches, which can be achieved using chiral dopants with chemical structures similar to those of the prototype ferroelectric nematics. The study of this type of mixtures is our plan for the near future.

Acknowledgements

This work was supported by the MICIU/AEI/FEDER, EU funds [project PGC2018-093761-B-C31] and the Gobierno de Aragón-FSE [E47_20R].